\documentclass[twocolumn,showpacs,prb,superscriptaddress]{revtex4}

\usepackage{graphicx}
\usepackage{dcolumn}
\usepackage{bm}

\newcommand*{\nhx}
{(NH$_3$)$_{x}$NaK$_2$C$_{60}$}

\begin{document}
\title{Metal-to-insulator evolution in (NH$_{3}$)$_{x}$NaK$_{2}$C$_{60}$: an
NMR study}
\author{M.~Ricc\`o}
\email{Mauro.Ricco@fis.unipr.it}
\homepage{http://www.fis.unipr.it/~ricco/}
\author{G.~Fumera}
\author{T.~Shiroka}
\author{O.~Ligabue}
\author{C.~Bucci}
\affiliation{Dipartimento di Fisica and Istituto Nazionale per la Fisica della Materia,\\
Universit\`a di Parma, Parco Area delle Scienze 7/a, 43100 Parma, Italy}
\author{F.~Bolzoni}
\affiliation{Istituto Maspec-CNR, Parco Area delle Scienze, Loc.\ Fontanini, 43010 Parma,
Italy}
\date{\today}

\begin{abstract}
A singular evolution toward an insulating phase, shown by $^{23}$Na and $%
^{13}$C NMR, has been observed in the superconducting fullerides (NH$_{3}$)$%
_{x}$NaK$_{2}$C$_{60}$ for $x>1$. Unlike most common cases, this insulating
phase is non magnetic and $^{13}$C spin lattice relaxation shows the
presence of a spin gap. These two features suggest that a charge
disproportion from C$_{60}^{3-}$ to C$_{60}^{2-}$ and C$_{60}^{4-}$ can
drive the system from the metallic to the insulating state.

The restoring of the Na$^{+}$ cation in the center of the octahedral
interstice in the insulating phase, as indicated by $^{23}$Na and $^{2}$H
lineshape analysis, confute the current belief that the cation off-centering
is effective in quenching the superconductivity.
\end{abstract}

\maketitle

%71.30.+h   Metal-insulator transitions
%74.70.Wz   Fullerenes and related materials
%74.20.Mn   Nonconventional superconductivity mechanisms (spin fluctuations, polarons and bipolarons)

\section{Introduction}

Soon after the discovery of fullerene it was found that compounds of the A$%
_3 $C$_{60}$ family, obtained by insertion of alkali metals (A) in the C$%
_{60}$ lattice, exhibited superconducting behaviour with high critical
temperatures $T_c$ (up to $\sim 30$ K). A regular increase in transition
temperatures was observed as the cation radius and lattice parameter
increased, a trend considered nowadays as a textbook example of the success
of standard BCS theory. Indeed, according to this theory, $T_c$ is strongly
dependent upon the density of states at Fermi level $\rho(E_{\text{F}}$), a
quantity which, as a rule, increases with the band-narrowing caused by
lattice expansion. This mechanism for increasing $T_c$ proved rather
successful, so attempts to find alternative ``lattice expanders'' soon
followed, thus suggesting the neutral molecule of ammonia as a good
candidate.

A well known example is given by Na$_{2}$CsC$_{60}$\cite{Zhou93}. After
ammoniation, it becomes (NH$_{3}$)$_{4}$Na$_{2}$CsC$_{60}$, with an increase
in transition temperature from 10.5 K to 29.6 K. In general, a further
narrowing of the band is expected to enhance the electron correlation, which
eventually induces a Mott-Hubbard transition to an insulating magnetic 
phase\cite{Han00}. This is just the case of K$_{3}$C$_{60}$, K$_{2}$RbC$_{60}$
and Rb$_{3}$C$_{60}$, which after 
ammoniation\cite{Iwasa96,Kitano02,Arvanitidis03,Takenobu00} become insulating 
and antiferromagnetic. In particular the superconductivity of 
NH$_{3}$K$_{3}$C$_{60}$ is recovered after application of external 
pressure\cite{Zhou95}.

An exception to this simple picture is given by the \nhx\ and 
(NH$_3$)$_{x}$NaRb$_2$C$_{60}$ compounds. Here the progressive \textit{increase} 
of ammonia concentration $x$ yields an \textit{increase} in lattice parameter
and an unexpected \textit{decrease} of $T_c$ \cite{Shimoda96}. This
anomalous behaviour was originally attributed to the octahedral cation
off-centering \cite{Shimoda96}. Indeed, in these systems the NH$_3$-Na
groups occupy the large octahedral (O) sites of the \textit{fcc} lattice and
the breaking of the cubic symmetry of the crystal field could lift the
degeneracy of the C$_{60}$ $t_{1u}$ LUMO, thus inducing a bandwidth increase
and a corresponding decrease in $\rho(E_{\text{F}})$. In a recent 
work\cite{Ricco01} we have found that in \nhx\ with $0.5<x<0.9$ the 
$\rho(E_{\text{F}})$, as extracted from Pauli spin susceptibility, actually 
increases on increasing $x$ and the lattice parameter, thus resulting 
\textit{inversely correlated} with $T_c$ (anti-Migdal behaviour). Accordingly, 
the superconducting critical parameters of the same system (upper and lower
critical fields $H_{c2}$, $H_{c1}$, and London penetration depth $\lambda$) 
\cite{Ricco03}, are significantly different from those of other
superconducting fullerides, and do not seem to fit the Ginzburg-Landau
theory predictions.

These facts clearly suggest that the role of ammonia in the superconducting
phase of \nhx\ ($x<1$) is quite different from that in the other fullerides,
with the detailed properties of its SC state also altered, showing features
which remind those of polaronic systems \cite{Ricco03}.

It is therefore necessary to investigate the properties of \nhx\ when larger
concentrations of ammonia are present and thus answer the questions of
whether or not the system evolves into an insulating phase, as the $T_c$
vs.\ $a$ behaviour would suggest, and, if so, whether the Na off-centering
has any role in the process.

This is the purpose of the present work in which we report the evolution of
the \nhx\ system as the doping increases beyond $x=1$. In what follows, a
detailed NMR study of $^{13}$C (spin lattice relaxation and lineshape) and 
$^{23}$Na (line shift) provide evidence for the formation of an 
insulating phase which segregates and increases in fraction as the 
ammonia concentration is increased. 
The electronic and structural properties of this new phase, in which the 
Na$^{+}$ ions recover the octahedral centered position, are also 
investigated by $^{23}$Na and $^{2}$H lineshape analysis.

\section{Sample Preparation and Characterization}

The \nhx\ samples were prepared following the procedure outlined in 
Ref.~\onlinecite{Shimoda96}. However, to achieve higher $x$ values, the final
vacuum drying from the ammonia solution was performed at lower temperatures
(down to $-23^{\circ }$C). The accurate ammonia concentration $x$ was
extracted from the hydrogen NMR line intensity. This procedure, besides
being quite reliable, was consistent also with volumetric methods which, on
the other hand, require the destruction of the sample.

Laboratory x-ray powder diffraction confirms the known \cite{Shimoda96} 
\textit{fcc} structure for the compounds with $x<1$ and, on the other hand,
it does not show any change in the structure on increasing ammonia doping,
apart from a slight increase in lattice parameter ($a$) from 14.39 \AA\ for 
$x=0.9$ to 14.44 \AA\ for $x=2$.

SQUID magnetometry shows a progressive decrease in $T_{c}$ from 15 K to 9.5
K for ammonia doping in the $0.5<x<1$ range. Above $x=1$ (corresponding to
the lowest transition temperature 9.5 K) up to $x=2$, no further decrease in 
$T_{c}$ takes place, but instead a \textit{progressive decrease of the
shielding fraction was observed}, consistent with the gradual formation of a
non-superconducting phase. No magnetism was detected in this range.

As mentioned in the introduction, a similar behavior has been already
observed in K$_{3}$C$_{60}$, K$_{2}$RbC$_{60}$ and Rb$_{3}$C$_{60}$ which,
upon ammoniation, show a Mott transition to an insulating magnetic 
phase\cite{Iwasa96,Kitano02,Arvanitidis03,Takenobu00}. In our case, however, 
the \textit{absence} of magnetism and the modest lattice expansion suggest that
other factors may play a role in suppressing the (super)conducting state.

\section{$^{23}$N\lowercase{a} NMR}

Nuclear Magnetic Resonance has proved a valuable tool for the study of the
electronic properties of fullerides since the early time of their discovery 
\cite{Tycko92,Tycko93}. Its usefulness mainly arises from the sizeable
hyperfine interaction of conduction electrons both with carbon and alkali
nuclei. In addition, microscopic structural information can be obtained also
from the nuclear quadrupole interactions with electric field gradients
(EFG). The latter interaction, however, affects only the resonance of
nuclear species with spin $I>1/2$, so the possible candidates in our case
are Na and K. We focused our attention on $^{23}$Na nucleus because, thanks
to its octahedral position: (i) a large hyperfine interaction with
conduction electrons is expected\cite{Rachdi97}, (ii) its off-centered
position implies the presence of a considerable electric field gradient on
it.

\begin{figure}[tbp]
\caption{$^{23}$Na NMR spectra of \nhx\ at RT. The sizeable Knight shift 
($\sim 165$ ppm) observed for $x<1$ disappears for $x\sim 2$ indicating 
a transition to an insulating phase.}
\label{fig:NaNMR}\includegraphics[width=0.45\textwidth]{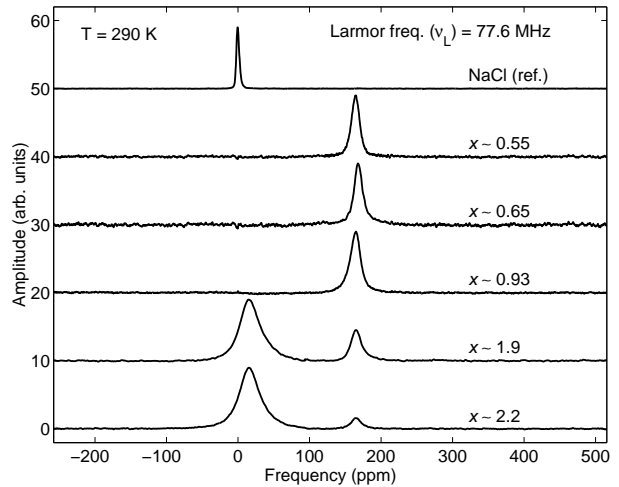}
\end{figure}

Figure~\ref{fig:NaNMR} shows the $^{23}$Na NMR spectra taken at RT in
samples with different ammonia contents in an external field of 6.88 T; the
measured resonances comprise all the three transitions, $^{23}$Na being an 
$I=3/2$ nucleus (see further).

%Figure 1
Let us first consider the \textit{shifts} of the lines with respect to a
reference NaCl acqueous solution. We notice that, as long as the samples are
superconducting, we measure a large paramagnetic shift (165 ppm), which is is 
attributed to the spin polarization of the conduction electrons (Knight shift). 
The shift and the shape of the resonance are consistent with an isotropic (Fermi)
hyperfine interaction as expected by the closed shell electronic structure
of the ion. The metallic character of the samples with $x<1$ is not a
surprise (this phase having been extensively investigated in previous 
works\cite{Shimoda96,Ricco01,Ricco03}). However, upon increasing the ammonia
nominal concentration above $x=1$, a peak at the reference $^{23}$NaCl
resonance frequency progressively appears and it becomes the dominant one
near $x=2$, representative of the $^{23}$Na behaviour in an insulating phase.

The residual paramagnetic shift, which is evident in the $x>1$ spectra of
Figure~\ref{fig:NaNMR}, could in principle suggest that the observed 150 ppm
shift may be attributed to a (big) change in hyperfine interaction with
conduction electrons from the $x<1$ phase to the $x=2$ phase and not to a
localization of the carriers induced by the developing of an insulating
phase. However, the same displacement of the $^{23}$Na line was observed
also in superconducting samples ($x<1$) which were exposed to a weakly
oxidizing atmosphere (O$_{2}\sim1000$ ppm). This treatment suppressed the
(super)conductivity of the sample without perturbing its structure (no
change in the diffraction pattern). This observation supports therefore the
suggestion that the disappearance of the paramagnetic shift is indicative of
an insulating phase and the residual shift can be attributed to a chemical
shift component. As we will see below, also $^{13}$C NMR results will
further confirm this issue.

It is important to notice that this system shows two different regimes with
respect to ammonia doping: (i) in the range $0.5<x<1$ it behaves as a
homogeneous solid solution, irrespective of the intrinsic disorder due to
the non-integer stoichiometry; this is guaranteed by the fact that
differently doped systems behave like homogeneous superconductors with a
well defined, $x$-dependent transition temperature; (ii) above $x=1$ the
observation of two coexisting phases (one superconducting and the other
insulating) indicates, on the other hand, that the separation of the two
differently doped phases is energetically favourable with respect to the
formation of a solid solution. The transition temperature of the residual
superconducting phase, observed for $x>1$, identifies its stoichiometry as 
$x=1$, while its disappearance as $x$ approaches 2 suggests the emergence 
of an insulating phase with $x=2$.

Laboratory x-ray powder diffraction shows only cubic reflections in the
whole range $0.5<x<2$, while a further increase of ammonia doping (above 
$x\simeq2$) results in the appearance of tetragonal reflections, which
indicate the formation of a different, ammonia richer phase. To avoid
contamination with this other (also insulating) phase, we performed our
measurements on the sample with $x=1.9$. We would like to stress that the
presence of a minority superconducting phase, evident also from 
Fig.~\ref{fig:NaNMR}, (and probably due to ammonia de-intercalation during 
the preparation/manipulation of the samples) does not affect the results
obtained in this work, since any possible contribution coming from this
phase in the measured spectra has been properly identified and taken into
account.

\begin{figure}[tbp]
\caption{Temperature dependence of the $^{23}$Na lineshape for metallic
(left) and insulating (right) \nhx\ samples. The large quadrupolar
broadening observed in the metallic phase ($\protect\nu _{\text{Q}}\sim 2$
MHz as estimated from the fit of the second order powder lineshape at low
temperature) is dramatically reduced ($\sim 1000$ times) in the insulating
phase.}
\label{fig:NaT}\includegraphics[width=0.45\textwidth]{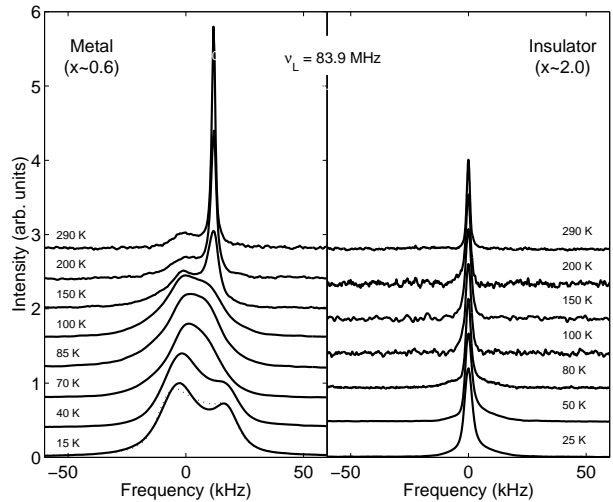}
\end{figure}

As mentioned before, the quadrupolar interaction of the Na nuclei with the
electric field gradient is expected to considerably broaden or split the
observed spectra, so that a comparison of the resonance \textit{line-widths}
and \textit{structure} turns out to be very illuminating for the
identification of Na$^{+}$ ion sites in the two phases. With reference to
the left side of Fig.~\ref{fig:NaT}, when $x<1$ the width at 290 K is
surprisingly small, if we recall that Na$^{+}$ ions are assumed to occupy
off-center positions in the octahedral interstices, where the local electric
field gradients are quite large. Even if we assume that the observed line
corresponds to the central $-\frac{1}{2}\leftrightarrow \frac{1}{2}$
transition, the expected second order quadrupolar broadening would be much
greater than the observed one \cite{Cohen57}. We attribute the observation
of a narrow line to a rotational narrowing process. Indeed, thanks to the
traceless nature of the quadrupolar interaction, an isotropic and fast
rotational motion would be effective in averaging it out. The broadening of
the line will be, in this case, restored only by freezing the rotational
dynamics upon lowering the sample temperature. The temperature dependence of
the $^{23}$Na resonance line in the two cases: $x=0.6$ and $x=2$
(superconducting and insulating phase respectively) is shown in 
Fig.~\ref{fig:NaT}.
%Figure 2

In the \textit{metallic phase} the expected broadening is progressively
reached on lowering the temperature. At the base temperature ($T=15$ K),
only the $-\frac{1}{2}\leftrightarrow \frac{1}{2}$ transition (affected only
at the second order by the quadrupolar interaction\cite{Cohen57}) survives,
while the other two ($-\frac{3}{2}\leftrightarrow -\frac{1}{2}$ and 
$\frac{1}{2}\leftrightarrow \frac{3}{2}$, directly affected by the quadrupolar
interaction) have moved too far in frequency to be observed. Further
evidence supports this interpretation: (i) the different lengths of the 90
degree pulse at high and low temperatures: a 50\% reduction is found in the
latter case, where only the central transition is observed\cite{Cohen57};
(ii) the lineshape at low temperature can be fitted quite well using a
second order quadrupolar broadening, as shown in the lowest spectrum of 
Fig.~\ref{fig:NaT}. A careful modelling of the observed lineshape at 
different applied fields\cite{Riccounpub} yields an estimate of 2.19 \AA\ 
for the off-centering along the [111] direction of the Na$^{+}$ ion, 
in complete agreement with recent diffraction studies\cite{Margadonna02}.

All these findings confirm that a motional narrowing process, rather
effective in averaging to zero the traceless quadrupolar interaction, takes
place at high temperatures. The small bump observed at reference frequency
in the room temperature spectrum is assigned instead to the central
transition of the remaining $(1-x)$ fraction of Na$^{+}$ ions, located in
tetrahedral positions. 
It appears unshifted due to the smaller hyperfine interaction of these 
ions\cite{Rachdi97}, whereas its rather large frequency spread could arise 
from the static quadrupolar interaction (even at room temperature) of the
tetrahedral Na$^{+}$ ions with the EFG created by the neighbouring C$_{60}$
units. 

Let's now turn to the \textit{insulating phase} (shown in the right side of
Fig.~\ref{fig:NaT}). In this case the $^{23}$Na resonance line is both
unshifted and unbroadened at all the investigated temperatures (only at $%
\sim 25$ K a small broadening appears, probably due to the slowing down of
the lattice vibrational dynamics and of C$_{60}$ rotation). If a quadrupolar
interaction were present, it would be small and it would induce a weak second
order broadening of the central $-\frac{1}{2}\leftrightarrow \frac{1}{2}$
line, in addition to possible satellite transitions. To ascertain such
quadrupolar effects, the detection bandwidth was extended to $\pm 500$ kHz
(a sufficient upper limit, considering the observed peak width). However, no
satellites were detected, so we conclude that the narrow resonance is due to
a vanishing quadrupolar interaction which merges all transitions in a single
resonance line.

To further support this conclusion we performed also an accurate measurement
of the 90 degree pulse length, which would be reduced by a factor $\frac{1}{2}$, 
if only the central transition were irradiated. The pulse length and power for
optimal excitation were found to be the same as in NaCl solution, where a single 
line comprises all the transitions.
Moreover, we checked if this condition was still valid when the power was
reduced and the pulse length extended, such as to irradiate only a narrow
band around the peak. Since no reduction of the effective 90 degree pulse
length, with respect to the NaCl solution, was noticed, we confirm that 
the observed resonance comprises all the transitions thus indicating a 
vanishing quadrupolar interaction.

The absence of any measurable electric field gradient, in spite of the 
large $^{23}$Na quadrupole moment 
($Q_{\text{Na}}\simeq0.11\times 10^{-28}$ m$^{2}$), is exactly what we 
expect in the centre of an environment with cubic symmetry.

In conclusion we find that: a) unlike the superconducting phase, in the
insulating one the Na$^{+}$ cation is located in the centre of the
octahedral site and, b) the \textit{fcc} structure evidenced by diffraction
is confirmed.

These findings rise some doubts on the current belief, which associates the
decrease in $T_{c}$ to the cation off-centering\cite{Shimoda96}. In fact,
according to that hypotesis, we would still find a superconducting phase at 
$x=2$ with a sizeable increase in $T_{c}$ instead of the observed insulator.

\section{$^{2}$H NMR}

\begin{figure}[tbp]
\includegraphics[width=0.45\textwidth]{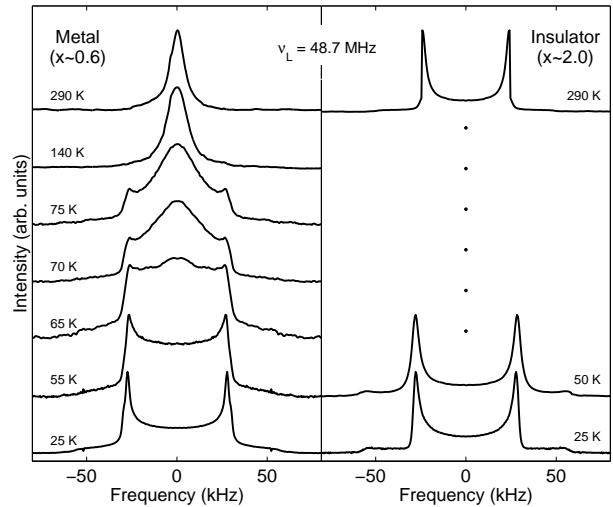}
\caption{$^2$H NMR spectra show typical Pake doublets at low temperatures
both in the metallic and in the insulating sample. The gradual motional
narrowing as the temperature rises, observed in the metallic sample, does
not occur in the insulating one because of the hampered NH$_3$ rotation in
it.}
\label{fig:Pake_doubl}
\end{figure}

As a further test and for a better understanding of the motional narrowing
mechanism evidenced by $^{23}$Na NMR, we extended our investigation to
samples prepared using deuterated ammonia, which allow to carry out $^2$H
NMR. Deuterium resonance ($I_{^2\text{H}}=1$) is dominated by the
quadrupolar interaction with the electric field gradient which, in this
case, originates from the $\sigma$ bond with the nitrogen\cite{Doverspike86}. 
Thus, the powder lineshape expected for a uniform distribution of ammonia
orientations is the typical Pake doublet\cite{Slichter90} shown in the
lower part of Fig.~\ref{fig:Pake_doubl}. 
%Figure 3
The expected width of 245 kHz (defined as the distance between
singularities) is however reduced by the uniaxial rotation of ammonia around
its ternary axis\cite{Doverspike86}, a mechanism we have found to be
effective down to 1.6 K. %% \cite{Ricco02b}. 
This dynamics is known\cite{Doverspike86} to produce an overall reduction of
the line width by a factor of $\sim 4.5$ without affecting the spectrum
shape. On the other hand, the reorientation of the ternary axis itself can
be quite efficient in averaging out the residual traceless quadrupolar
interaction, thus making the powder pattern collapse into a single line, as
shown in the upper part of Fig.~\ref{fig:Pake_doubl} for $x=0.6$
(superconducting phase), while no narrowing was observed for $x=2$
(insulating phase).

These results can be rationalized as follows: in the superconducting phase
at high temperatures the NH$_3$-Na clusters rotate around the centre of the
octahedron thus averaging to zero the quadrupolar interactions of both $%
^{23} $Na and $^2$H. Upon cooling down from RT this dynamics slows down and
the situation appears static at $T \sim 55$ K on the NMR timescale, i.e.\
the correlation times are expected to be larger than 0.2 ms. The presence of
two ammonia molecules in the insulator ($x=2$ case) inhibits these
rotations, which appear to be frozen already at RT.

\begin{figure}[tbp]
\includegraphics[width=0.45\textwidth]{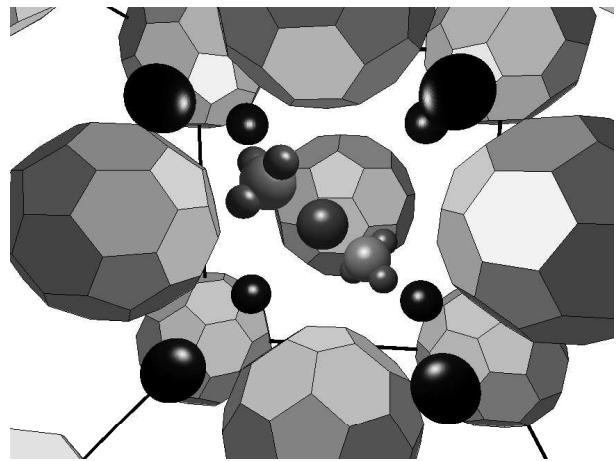}
\caption{Pictorial view of the (NH$_{3}$)$_{2}$NaK$_{2}$C$_{60}$ structure
as inferred from NMR and x-ray measurements. For $x=2$ the Na$^{+}$ ion is
\textquotedblleft sandwiched\textquotedblright\ between two ammonia
molecules at the centre of the octahedral interstice.}
\label{fig:offcenter}
\end{figure}

Since the $^{23}$Na NMR suggests that the Na$^{+}$ ion is restored in the
centre of the octahedral site, we can envisage the structure of the
insulating phase as that represented in Fig.~\ref{fig:offcenter}.
%Figure 4
%
% FOR ONLINE COLOR VERSION
%
In spite of the very modest lattice expansion, the O site can still easily
host the NH$_{3}$-Na-NH$_{3}$ cluster. By comparison with 
(NH$_{3}$)$_{4}$Na$_{2}$CsC$_{60}$ \cite{Zhou93}, and by assuming the 
same Na-N distance (2.5 \AA ), only a 1.5\% reduction (from 2.29 to 
2.26 \AA ) of the closest H-C distance takes place.

\section{$^{13}$C NMR}

\begin{figure}[tbp]
\includegraphics[width=0.45\textwidth]{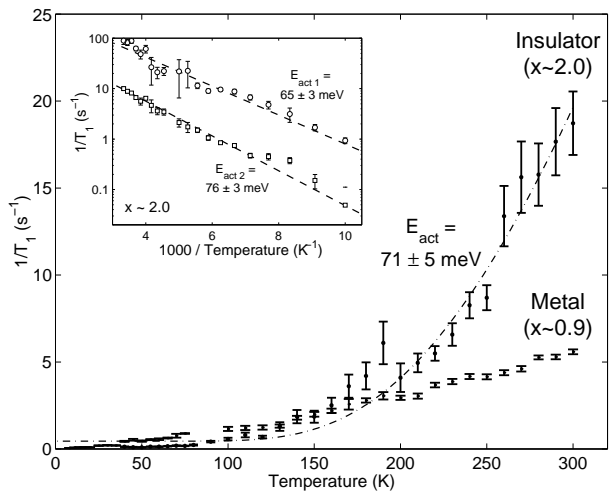}
\caption{$^{13}$C NMR $1/T_{1}$ vs.\ temperature dependence in insulating
and metallic (NH$_{3}$)$_{x}$NaK$_{2}$C$_{60}$. Whereas the latter follows a
linear Korringa law, the insulator shows two thermally activated exponential
relaxations, as reported in the inset. The fit with a single relaxation
shown in the main figure (doted line) is characterized by an average 
$E_{\text{act}}=71\pm 5$ meV spin-gap.}
\label{fig:T1relax}
\end{figure}

The lack of a magnetic response in the insulating phase, as evidenced by
SQUID measurements, suggests that this phase does not arise from a classical
Mott transition as the one observed, for example, in NH$_{3}$K$_{3}$C$_{60}$ 
\cite{Tou00} (where an antiferromagnetic phase exists below $T_{N}=40$ K 
\cite{Iwasa96}). To tackle this problem and to confirm the previous findings
we consider a third nuclear probe available in our system: the $^{13}$C
nucleus, whose spin-lattice relaxation time ($T_{1}$) both in the
superconducting and in the insulating phase can yield information on
electron spin and/or density fluctuations. Figure~\ref{fig:T1relax} shows
the spin-lattice relaxation rate $1/T_{1}$ as a function of temperature. 
%Figure 5
While a linear behaviour is observed in the $x<1$ phase, as expected on the
basis of the Korringa relation for metals \cite{Slichter90}, the $x\sim2$
phase follows an activated law, as shown in the fit of Fig.~\ref{fig:T1relax}, 
where $E_{\text{act}}=71\pm 5$ meV. The disappearance of the Korringa
behaviour in this latter phase and its lower relaxation rate at low
temperatures (0.1 s$^{-1}$ for $x\sim2$ at 45 K to be compared with 0.55 
s$^{-1}$ for $x\sim0.9$), both indicate the absence of interactions with
conduction electrons in the relaxation process, thus further confirming its
insulating nature. By a comparison of these results with the $^{13}$C
relaxation data in similar insulating systems (i.e.\ Na$_{2}$C$_{60}$ and 
K$_{4}$C$_{60}$)\cite{Brouet01}, we attribute the activated behaviour to 
the interaction with a thermally excited triplet state. The reorientational
dynamics of C$_{60}$ can also contribute, through the modulation of the
chemical shift anisotropy (CSA), to the relaxation of carbon nuclei, but the
evaluation of this term\cite{Zimmer95,Brouet01} and the comparison with
other known cases (for example Na$_{2}$C$_{60}$\cite{Brouet00}) show that
the expected contribution is restricted to a bump in 1/T$_{1}$ located
around $T=180$ K, visible also in our case. The interaction with excited
electrons is therefore the only responsible for the observed activated
behaviour of the relaxation, thus suggesting the presence of an electron
energy gap.

Moreover, a careful analysis of the recovery curves shows a \textit{double
exponential behaviour} (the amplitude of the differently relaxing components
being almost equal), which is more pronounced in the high-temperature region
(100--300 K). Although a non-exponential relaxation could be ascribed also
to the inequivalence of different carbons in the \textit{fcc} lattice or to
fluctuations of anisotropic interactions (i.e.\ CSA or Knight 
shift)\cite{Mehring94}, in both these cases the effect would be more pronounced 
at low temperatures. Since the opposite is observed, the double exponential
relaxation is more likely due to the presence of two classes of
non-equivalent carbon nuclei, both relaxing through activated electronic
spin-triplet states. In this case we expect different relaxation rates
throughout the temperature range. 
The double exponential recovery thus indicates the presence of two spin gaps
with different activation energies. The inset of Fig.~\ref{fig:T1relax}
shows the temperature behaviour of the two components in the recovery curve.
Fitting the data with the function $1/T_{1}=Ae^{-\frac{E_{a}}{k_{\text{B}}T}%
}+\text{const.}$ yields respectively $A_{1}=914\pm 84$ s$^{-1}$, $%
E_{a1}=65\pm 3$ meV and $A_{2}=185\pm 11$ s$^{-1}$, $E_{a2}=76\pm 3$ meV.

The presence of a singlet-triplet gap\cite{Brouet01}, like the one observed
here, is typical of Jahn-Teller (JT) distorted C$_{60}^{(2,4)-}$. This,
together with the lack of magnetism and the presence of two different energy
gaps, indicates that a quasi-static charge unbalance or, in chemical
language, a disproportion as $\text{C}_{60}^{3-}\rightarrow \text{C}%
_{60}^{2-}+\text{C}_{60}^{4-}$ could take place in our system. The different
pre-exponential factors could account for two different hyperfine
interactions of the paramagnetic electrons with the $^{13}$C nuclei. Their
ratio ($A_{1}/A_{2}\sim5$) is close to the value ($\sim4$) reported for Na$%
_{2}$C$_{60}$ and K$_{4}$C$_{60}$\cite{Brouet01}. The two activation
energies, on the other hand, are quite close, in agreement with theoretical
predictions\cite{Manini94}. It is interesting to note that recent studies by
V.\ Brouet et al., also based on NMR relaxation measurements, report the
observation of a similar JT induced spin gap not only in C$_{60}^{(2,4)-}$
systems\cite{Brouet01}, but even in conducting CsC$_{60}$\cite{Brouet02} and
superconducting Na$_{2}$CsC$_{60}$ and Rb$_{3}$C$_{60}$\cite{Brouet03}, in
which the presence of C$_{60}$ anions with even charges is explained as the
result of fast dynamical charge disproportions or fluctuations. If we
suppose that the same effect could be present in the metallic phase of \nhx\
(although it could not be observed in relaxation time measurements for other
reasons\cite{Brouet03}), the known anomalies of this superconductor and the
observed evolution to the insulating phase could be the effect of the
slowing down, and eventually of a freezing, of these charge fluctuations.

The $^{13}$C\ spectra taken at low temperature ($T=20$ K), where narrowing
due to molecular reorientational dynamics is ineffective, show in both cases
Gaussian lineshapes. They are roughly centered at 200 ppm with respect to
TMS (tetramethylsilane) and have different FWHM values: 118 ppm for $x\simeq
0.9$ and 177 ppm for $x\simeq 2.0$. Both the linewidths and their positions
fully agree with previous observations: those for $x\simeq 0.9$ doping
conform to the values found in other superconducting fullerides\cite{Sato98}, 
whereas the values for $x\sim 2$ doping agree with those measured in the C$%
_{60}^{2-}$ and C$_{60}^{4-}$ insulating systems\cite{Brouet00,Zimmer94}.
The linewidth observed in the metallic phase, smaller than that in the
insulator, can be attributed to the different sign of the CSA with respect
to the Knight shift anisotropy, yielding a line narrowing once the two
tensors are added. These observations are also consistent with the
insulating nature of the $x=2.0$ phase. 

\section{Conclusions}

In conclusion we have shown that \nhx\ for $x>1$ evolves toward a rather
singular insulating phase which, unlike other known cases, does not show any
magnetic order. Moreover, this phase is hardly imputable to a lattice
expansion but, more likely, it originates from a charge disproportion among 
C$_{60}$ anions, evidenced by the presence of a spin gap. According to
theoretical analysis \cite{Fabrizio97}, in odd electron systems JT effect
adds to the Coulomb repulsion to localize electrons in a non magnetic ground
state, while the opposite happens in even electron systems. The related gain
in energy could drive our system to favour a JT induced transition towards
the insulating state rather then following the conventional Mott-Hubbard
route. This system would represent the first example of a JT induced
metal-to-insulator transition in superconducting fullerides.

Similar charge instabilities could also be at the origin of the anomalous 
$T_{c}$ vs.\ lattice parameter observed in the (super)conducting phase of our
system. This suggestion is supported by the recent observation of a similar
spin gap in the conducting CsC$_{60}$\cite{Brouet02} and superconducting 
Na$_{2}$CsC$_{60}$ and Rb$_{3}$C$_{60}$\cite{Brouet03}, where a fast 
dynamical disproportion might be detectable already in the metallic phase.

Fundamental questions remain however open to deeper theoretical and
experimental investigations, regarding the compatibility between the charge
disproportion mechanism on one side and the non conventional BCS nature of
the superconducting condensate on the other.

This non conventional behaviour is clearly evidenced in the superconducting
phase of the systems studied in the present work and confirmed also by
recent experiments involving the density of states at the Fermi level 
\cite{Ricco01} or the critical magnetic fields and the field penetration 
depth\cite{Ricco03}.

\bibliography{nh3_ricco}

\end{document}